\documentclass[showpacs,aps,twocolumn]{revtex4}
\usepackage{bm}
\usepackage{amsmath}
\usepackage{graphicx}
\usepackage[usenames,dvipsnames]{color}
\definecolor{darkblue}{RGB}{0,0,196}
\usepackage[colorlinks=true,linkcolor=darkblue,citecolor=darkblue,urlcolor=darkblue]{hyperref}

\usepackage{setspace}

\def\be{\begin{equation}}
\def\ee{\end{equation}}
\def\ba{\begin{eqnarray}}
\def\ea{\end{eqnarray}}

\def \l{\left(}
\def \r{\right)}
\def \l{\left(}
\def \r{\right)}

\usepackage{graphicx}
\usepackage{amsmath,bbm}
\usepackage{amssymb,bm}

\newcommand\la{\langle}
\newcommand\ra{\rangle}

\begin{document}

\title{Effect of Hagedorn States on Isothermal Compressibility of Hadronic Matter formed in Heavy-Ion Collisions: From NICA to LHC Energies}
\author{Arvind~Khuntia}
\author{Swatantra~Kumar~Tiwari}
\author{Pramod~Sharma}
\author{Raghunath~Sahoo\footnote{Corresponding author: $Raghunath.Sahoo@cern.ch$}}
\affiliation{Discipline of Physics, School of Basic Sciences, Indian Institute of Technology Indore, Simrol- 453552, Indore, INDIA}
\author{Tapan K. Nayak}
\affiliation{National Institute of Science Education and Research, Homi Bhabha National Institute, Jatni- 752050, Odisha, India} 
 \affiliation{CERN, CH 1211, Geneva 23, Switzerland}

\begin{abstract}

\noindent
In this work, we have studied the isothermal compressibility ($\kappa_T$) as a function of temperature, baryon chemical potential and centre-of-mass energy ($\sqrt{s_{NN}}$) using hadron resonance gas (HRG) and excluded-volume hadron resonance gas (EV-HRG) models. A mass cut-off dependence of isothermal compressibility has been studied for a physical resonance gas. Further, we study the effect of heavier resonances ($>$ 2 GeV) on the isothermal compressibility by considering the Hagedorn mass spectrum, ${\rho}(m)\sim{\exp(bm)}/{(m^2+m_0^2)^{5/4}}$. Here, the parameters, $b$ and $m_0$ are extracted after comparing the results of recent lattice QCD simulations at finite baryonic chemical potential. We find a significant difference between the results obtained in EV-HRG and HRG models at higher temperatures. The inclusion of the Hagedorn mass spectrum in the partition function for hadron gas has a large effect at a higher temperature. A higher mass cut-off in the Hagedorn mass spectrum takes the isothermal compressibility to a minimum value, which occurs near the Hagedorn temperature ($T_H$).  
We show explicitly that at the future low energy accelerator facilities like FAIR (CBM), Darmstadt and NICA, Dubna the created matter would have higher compressibility compared to the high energy facilities like RHIC and LHC.

\end{abstract}
\date{\today}
\maketitle 
 
\section{Introduction}
\label{intro}
Ultra-relativistic heavy-ion colliders such as Relativistic Heavy Ion Collider (RHIC) and Large Hadron Collider (LHC) aim to produce matter at extreme conditions of temperature and/or energy density, where a phase transition is expected to take place from colorless hadronic matter to a colored phase of quarks and gluons known as Quark-Gluon Plasma (QGP). Lattice QCD (lQCD) predicts a smooth crossover transition between hadron gas (HG) phase and QGP phase at zero baryon chemical potential~\cite{Aoki:2006we,Bazavov:2014pvz}. However, various QCD inspired phenomenological models predict a first order phase transition from HG to QGP phase~\cite{Gottlieb:1985ug,Fukugita:1989yb,Schaefer:2004en,Herpay:2005yr}. These results advocate the possible presence of a critical point (CP), where the first order phase transition ends. Currently, theoretical and experimental studies are dedicated towards understanding the nature of the QCD phase transition and location of CP. Thermodynamic observables such as specific heat and isothermal compressibility are very useful in quantifying the nature of the phase transition and to obtain the equation of state (EOS) of the matter. Specific heat measures the change in energy density with respect to the change in temperature and is also related to the temperature fluctuation in the system~\cite{Stodolsky:1995ds,Shuryak:1997yj,Basu:2016ibk}. The isothermal compressibility ($\kappa_T$) defines the change in volume with the change in pressure at constant temperature. It can be expressed as a second order derivative of Gibbs free energy with respect to pressure and is expected to diverge at CP. Thus, $\kappa_T$ and $C_V$ are very helpful in unveiling the nature of the phase transition and helpful in the search for CP.           
 
Various methods are employed to study $\kappa_T$ of hadrons produced in heavy-ion collision experiments. Ideal Hadron resonance gas (HRG) model is a type of statistical-thermal model, which is used to describe the behaviour and properties of hadrons in equilibrium~\cite{Andronic:2005yp,Cleymans:1998fq,Cleymans:2000ck,BraunMunzinger:2001ip,BraunMunzinger:1994xr,Cleymans:1997sw,Cleymans:1998yb,Becattini:1997ii,Torrieri:2001ue,Torrieri:2004zz,Wheaton:2004qb,Becattini:2000jw}. In this model, the hadrons and their resonances are assumed as ideal or non-interacting particles. HRG model is very successful in describing the particle ratios measured in various heavy-ion collision experiments~\cite{Andronic:2005yp,Cleymans:2000ck,BraunMunzinger:2001ip,BraunMunzinger:1994xr,Cleymans:1997sw,Cleymans:1998yb,Becattini:1997ii} as well as in nucleon-nucleus and nucleon-nucleon collisions~\cite{Sharma:2018uma,Sharma:2018owb}. However, when HRG is used to study the phase transition from hadron gas (HG) to quark-gluon plasma (QGP) phase using Gibbs construction, an anomalous reversal of phase from QGP to HG is observed at large baryon chemical potential and/or temperature~\cite{Tiwari:2012}. This ambiguity is removed by giving a hard-core size to each baryon which results in strong repulsive interactions between a pair of baryons or antibaryons. Such type of statistical-thermal model is known as excluded- volume hadron resonance gas (EV-HRG) model, which is also very successful in describing various properties of hadrons particularly at lower collision energies~\cite{Tiwari:2012,Cleymans:1986yj,Cleymans:1990ni,Davidson:1991um,Kouno:1988bi,Hagedorn:1980kb,Rischke:1991ke,Singh:1991np,Panda:2002iu,Anchishkin:1995np,Prasad:2000sg,Tiwari:1998jy}. Hagedorn has proposed an exponentially increasing continuous mass spectrum to describe a variety of particles called as Hagedorn mass spectrum. This is given as~\cite{Cleymans:2011fx}, 
\begin{eqnarray}
\label{eq1}
\rho(m)=\frac{A}{(m^2+m_0^2)^{5/4}}\exp(m/T_{H}). 
\end{eqnarray}
Here, $m_0$ and $A$ are the free parameters extracted by comparing the lattice QCD (lQCD) results with the hadron resonance gas (HRG) model including Hagedorn mass spectrum at finite baryochemical potential. The slope of this spectrum is defined by the limiting Hagedorn temperature, $T_H$. The partition function of a hadronic matter diverges above $T_H$ and a new phase is possibly formed called as quark-gluon plasma (QGP). Thus, in the infinite mass limit, the thermodynamical observables for the Hagedorn resonance gas would show critical behaviour as it reaches to $T_H$ ($\sim$150 $\pm$ 15 MeV~\cite{Rafelski:2016hnq}) which approximately equals to the critical temperature ($T_C\sim$ 159 $\pm$ 10)~\cite{Bazavov:2014pvz}. The thermodynamic and transport properties of hadrons modify significantly after incorporating the Hagedorn states (HS) in the HRG model. The speed of sound ($c_s$) becomes lower at the phase transition when one adds mass spectrum in HRG and describes the lQCD data successfully~\cite{NoronhaHostler:2008ju,Majumder:2010ik,NoronhaHostler:2012ug,Jakovac:2013iua,Khuntia:2016ikm}. A similar decrease is also found in shear viscosity to entropy ratio ($\eta/s$) when comparing the results obtained with HS~\cite{NoronhaHostler:2008ju,NoronhaHostler:2012ug,Gorenstein:2007mw,Itakura:2008qv}. These findings prompt us to carry out a study of effect of Hagedorn mass spectrum on isothermal compressibility ($\kappa_T$) of a hadron gas at very high temperature.

In this paper, we have presented the calculation of the isothermal compressibility of a hadronic matter produced in high energy collisions using hadron resonance gas and excluded-volume model. We study $\kappa_T$ as a function of temperature, baryon chemical potential, mass cut-off and centre-of-mass energy. We also see the behaviour of $\kappa_T$ after including the Hagedorn mass spectrum in the grand partition function for hadrons
in HRG. The experimental data~\cite{Mukherjee:2017elm} for isothermal compressibility at the Relativistic Heavy-Ion Collider (RHIC) at Brookhaven National Laboratory (BNL) and Super Proton Synchrotron (SPS) of CERN are also shown for comparison. The results of event generators such as A Multi Phase Transport (AMPT) model, Ultra-relativistic Quantum Molecular Dynamical (UrQMD) simulation and EPOS are presented for completeness~\cite{Mukherjee:2017elm}. 

The paper is organised as follows: in Section~\ref{Method}, we give the formulation of isothermal compressibility for a physical hadrons gas using HRG and EV-HRG models. Here, we have taken all the hadrons and their resonances having masses upto a cut-off value of 2 GeV. Further, we include those resonances which have well-defined masses and widths. The branching ratios for sequential decays are also accounted for in both the models. In addition, we present the calculation of $\kappa_T$ in HRG with Hagedorn mass spectrum. In Section~\ref{result}, we discuss the results obtained using this formulation. Finally, we summarise and conclude in Section~\ref{summary}.

\noindent
\section{Methodology}
\label{Method}
In this section, we present the formulations used to calculate the isothermal compressibility ($\kappa_T$) for a hot hadronic matter. In this work, first, we discuss hadron resonance gas (HRG) model used for calculation of $\kappa_T$. We then present the formulation of Excluded-Volume Hadron Resonance Gas model based on van der Waals type of interactions. Finally, we derive isothermal compressibility after including the Hagedorn mass spectrum in the HRG model. 

\subsection{Hadron Resonance Gas (HRG) Model}

In this subsection, we consider a non-interacting physical hadron resonance gas. The grand canonical partition function for a ideal system of $i^{\rm th}$ particle species contained in a volume $V$ in Boltzmann's approximation with mass $m_i$ and chemical potential $\mu_i$ is given as~\cite{Andronic:2005yp},

\ba
\label{eq3}
\ln Z_i = \frac{g_iV}{2 \pi^2}\int p^2dp~ \exp~[-(\varepsilon_i-\mu_i)/T],
\ea
where $g_{i}$ is the degeneracy factor and $\varepsilon_i=\sqrt{p^2+m_i^2}$ is the energy of $i^{\rm th}$ particle. The number density of hadrons calculated by using the basic thermodynamical relationship given as follows,

\ba
\label{eq4}
n_i = \frac{g_i}{2\pi^2}\int p^2dp~ \exp [-(\varepsilon_i-\mu)/T].
\ea
The isothermal compressibility is a measure of change of volume with the change in pressure at a constant temperature and average number of particles, which is given as~\cite{Mrowczynski:1997kz},

\ba
\left.\kappa_T\right|_{T,\la N_i\ra} &=&
-\frac{1}{V}\left.\l\frac{\partial V}{\partial P}\r\right|_{T,\la N_i\ra}
\label{eq5}
\ea
In order to calculate $\kappa_T$, we proceed as follows. The change in pressure can be written as,
\ba
dP = \l\frac{\partial P}{\partial T}\r dT + \sum_i\l\frac{\partial
  P}{\partial \mu_i}\r d\mu_i\label{eq6}, 
\ea
which gives, 
\ba
\left.\l\frac{\partial P}{\partial V}\r\right|_{T,\{\la N_i\ra\}} = \sum_i\l\frac{\partial P}{\partial \mu_i}\r
\left.\l\frac{\partial \mu_i}{\partial V}\r\right|_{T,\{\la N_i\ra\}}\label{eq7}.
\ea
The first factor is the formula for $N_i$ and the second factor is obtained from the condition of constancy of $N_i$ as follows,
\ba
dN_i = \l\frac{\partial N_i}{\partial T}\r dT + \l\frac{\partial N_i}{\partial V}\r dV + 
\l\frac{\partial N_i}{\partial \mu_i}\r d\mu_i\label{eq8}.
\ea

If $N_i$ and T are kept constant, then Eq.~\ref{eq8} reduces to
\ba
\left.\l\frac{\partial \mu_i}{\partial V}\r\right|_{T,\{\la N_i\ra\}}  = -\frac{\l\frac{\partial N_i}{\partial V}\r}
{\l\frac{\partial N_i}{\partial \mu_i}\r}\label{eq9}.
\ea
In the HRG model, $\displaystyle \frac{\partial N}{\partial V}=\frac{\partial P}{\partial\mu}$. Thus, Eq.~\ref{eq7} becomes,
\ba
\left.\l\frac{\partial P}{\partial V}\r\right|_{T,\{\la N_i\ra\}} 
= -\sum_i\frac{\l\frac{\partial P}{\partial \mu_i}\r^2}{\l\frac{\partial N_i}{\partial \mu_i}\r}\label{eq10}
\ea
Using Eqs.~\ref{eq5} and~\ref{eq10}, we get,

\ba 
\left.k_T\right|_{T,\{\la N_i\ra\}} =
\frac{1}{\sum_i\Big[{\l\frac{\partial P}{\partial \mu_i}\r^2}/{\l\frac{\partial n_i}{\partial \mu_i}\r}\Big]}\label{eq11}. 
\ea
Here, $n_i = N_i/V = \frac{\partial P}{\partial \mu_i}$ and calculated by using equation~\ref{eq4}.
 
\subsection{Excluded-Volume Hadron Resonance Gas (EV-HRG) Model}
Now, we describe an equation of state for the HG based on the excluded-volume correction~\cite{Tiwari:2012}, where a hard-core size is assigned to all the baryons while mesons are treated as point-like particles in the grand canonical partition function. An excluded-volume correction in physical hadron resonance gas indirectly takes care of van der Waals-like repulsive interactions that arise due to a hard core size of the particles. In this approach, an annihilation of baryons and anti-baryons is not allowed. The excluded number density is calculated as~\cite{Tiwari:2012},
\begin{equation}
n_i^{ex} = \frac{\lambda_i}{V}\left(\frac{\partial{ln Z_i^{ex}}}{\partial{\lambda_i}}\right)_{T,V}\label{eq12},
\end{equation}
with
\begin{equation}
ln Z_i^{ex} = V(1-\sum_jn_j^{ex}V_j^0)I_{i}\lambda_{i},
\end{equation}

Here, $V_j^0$ is the eigen-volume assigned to each baryon of $\rm{j^{th}}$ species and hence $\sum_{j}n_jV_j^0$ becomes the total occupied volume where $n_{j}$ represents the total number of baryons of $\rm j^{th}$ species.

Equation~\ref{eq12} leads to a transcendental equation:
\begin{equation}
n_i^{ex} = (1-R)I_i\lambda_i-I_i\lambda_i^2\frac{\partial{R}}{\partial{\lambda_i}}+\lambda_i^2(1-R)I_i^{'}\label{eq13},
\end{equation}
where $I_{i}^{'}$ is the partial derivative of $I_{i}$ with respect to $\lambda_{i}$ and $R=\sum_in_i^{ex}V_i^0$ is the fractional occupied volume. We can write R in an operator equation form as follows~\cite{Tiwari:2013wga}:
\begin{equation}
R=R_{1}+\hat{\Omega} R\label{eq14},
\end{equation}
where $R_{1}=\frac{R^0}{1+R^0}$ with $R^0 = \sum n_i^0V_i^0 + \sum I_i^{'}V_i^0\lambda_i^2$; $n_i^0$ is the density of point-like baryons of $i^{\rm th}$ species and the operator $\hat{\Omega}$ has the form:
\begin{equation}
\hat{\Omega} = -\frac{1}{1+R^0}\sum_i n_i^0V_i^0\lambda_i\frac{\partial}{\partial{\lambda_i}}\label{eq15}.
\end{equation}
Using Neumann iteration method and retaining the series upto $\hat{\Omega}^2$ term, we get
\begin{equation}
R=R_{1}+\hat{\Omega}R_{1} +\hat{\Omega}^{2}R_{1}\label{eq16}.
\end{equation}
\noindent
Once we get $R$, we can calculate the excluded number density and isothermal compressibility by using equations~\ref{eq11} and \ref{eq13}, respectively.
  
\subsection{Hagedorn Mass Spectrum}
In this subsection, we include exponentially increasing continuous mass spectrum of the Hagedorn form given by eq.~\ref{eq1} in the grand canonical partition function for a hadron gas, which is given as,
\ba
\ln Z = \frac{1}{2\pi^2} \int \exp\Big(-(\varepsilon - \mu)/T\Big)p^2 dp \nonumber \\
                              \times \int_{m_0}^\infty \rho(m) dm,
\label{eq17}
\ea
where, the spectral function, $\rho(m)$ can be written as,
\ba
\rho(m) = \frac{A}{(m^2+m_0^2)^{5/4}}\exp(m/T_H).
 \label{eq18}
\ea
Now, eq.~\ref{eq17} becomes,
\ba 
\ln Z =  \frac{1}{2\pi^2} \int \exp\Big(-(\varepsilon-\mu)/T\Big)p^2 dp \nonumber \\
                      \times \int_{m_0}^\infty \frac{A}{(m^2+m_0^2)^{5/4}}\exp(m/T_H)dm.
\label{eq19}
\ea
The values of $A$ and $m_0$, those arise from fitting the lQCD data are 0.57 GeV$^{-3/2}$ and 0.425 GeV, respectively~\cite{Lo:2015cca}. Here, the Hagedorn temperature, $T_H$ = 0.180 GeV~\cite{Lo:2015cca}. The formula for the number density and pressure of hadrons after including the Hagedorn mass spectrum is given as follows,

\ba
n = \frac{1}{2\pi^2} \int \exp\Big(-(\varepsilon-\mu)/T\Big)p^2 dp \nonumber \\
                      \times \int_{m_0}^\infty \frac{A}{(m^2+m_0^2)^{5/4}}\exp(m/T_H)dm,
\label{eq20}
\ea

and,

\ba
P = \int  \exp\Big(-(\varepsilon-\mu)/T\Big)~\frac{p^4}{3\varepsilon}dp  \nonumber\\  
        \times \int_{m_0}^\infty \frac{A}{(m^2+m_0^2)^{5/4}}\exp(m/T_H)dm.
\label{eq21}
\ea

The isothermal compressibility of hadrons in the presence of Hagedorn mass spectrum is calculated by using eq.~\ref{eq11}.

\begin{figure}[ht!]
\includegraphics[width=0.45\textwidth]{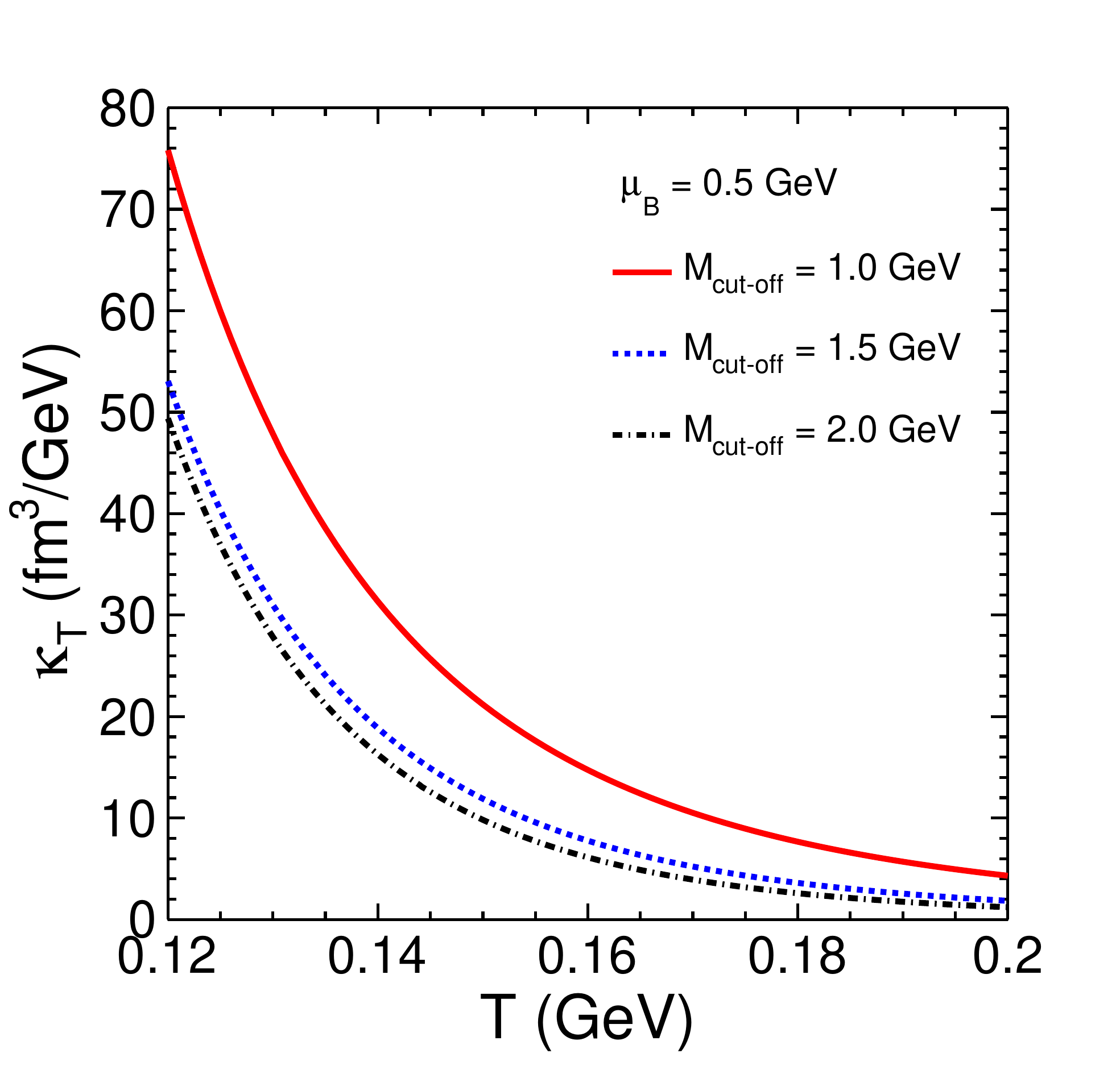}
\caption{(color online) The isothermal compressibilities obtained in HRG model as a function of temperature for various mass cut-offs at $\mu_B$ = 0.5 GeV.}
\label{kTmass}       
\end{figure}

\section{Results and Discussion}
\label{result}
In figure~\ref{kTmass}, we have shown the $\kappa_T$ as a function of temperature for various mass cut-offs at $\mu_B$ = 0.5 GeV calculated in HRG model. $\kappa_T$ decreases with temperature due to increase of number of particles. We also observe that isothermal compressibility has a lower value when higher mass resonances added into the system. This suggests that with more number of particles, it is difficult to compress the system and hence it becomes more incompressible. This could be because of particles gaining higher kinetic energy. Further, we have explicitly calculated the change in number density in HRG model with $\mu_B$ = 0.5 GeV at $T$ = 0.170 GeV for various mass cut-offs. We find 23\% increase in number density when mass cut-off increases from 1.0 GeV to 1.5 GeV.   
 
\begin{figure}[ht!]
\includegraphics[width=0.45\textwidth]{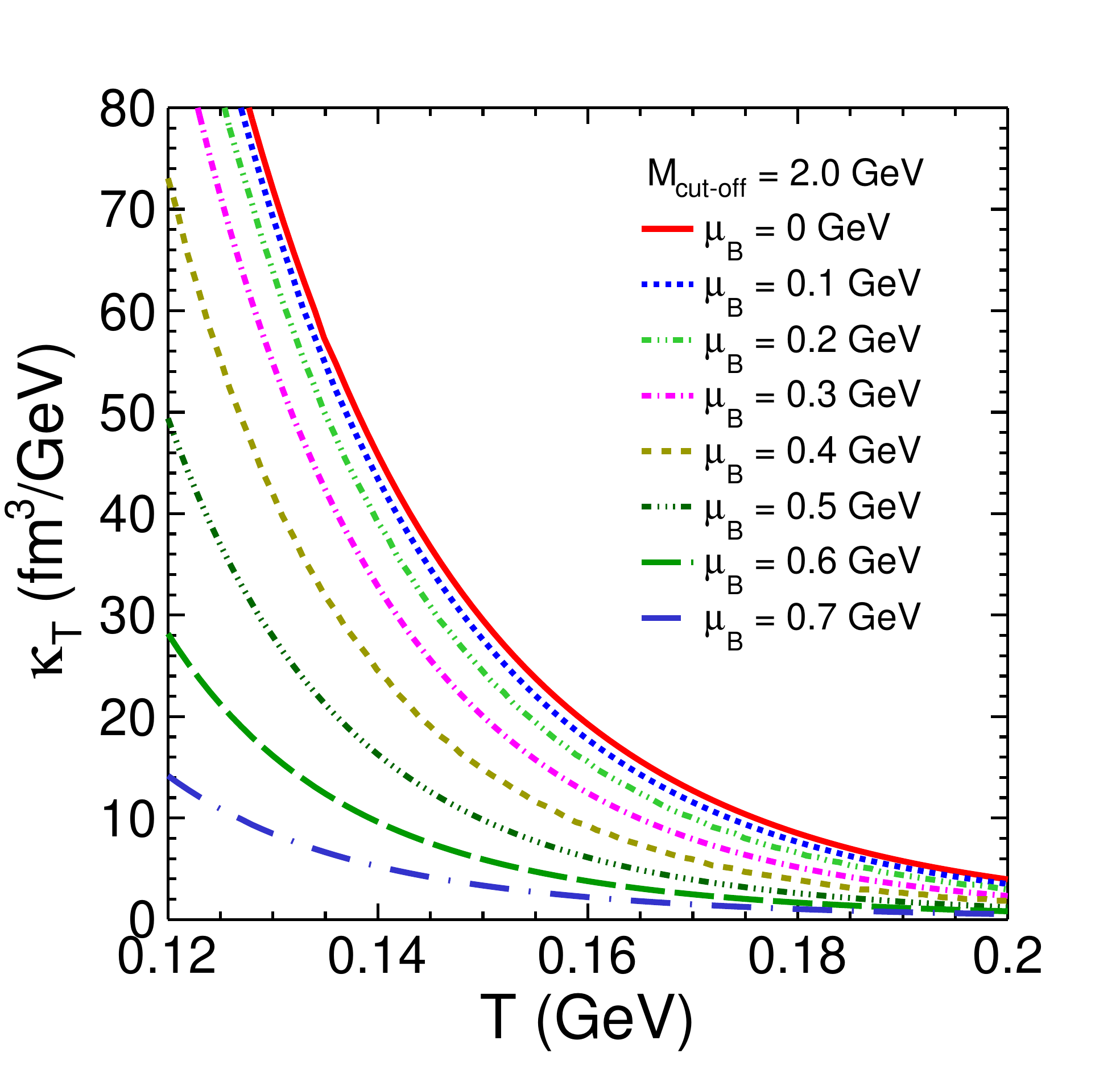}
\caption{(color online) $\kappa_T$ as a function of temperature for various $\mu_B$. The lines are the results obtained in HRG model for different $\mu_B$.}
\label{kTmu}       
\end{figure}

\begin{figure}[ht!]
\includegraphics[width=0.45\textwidth]{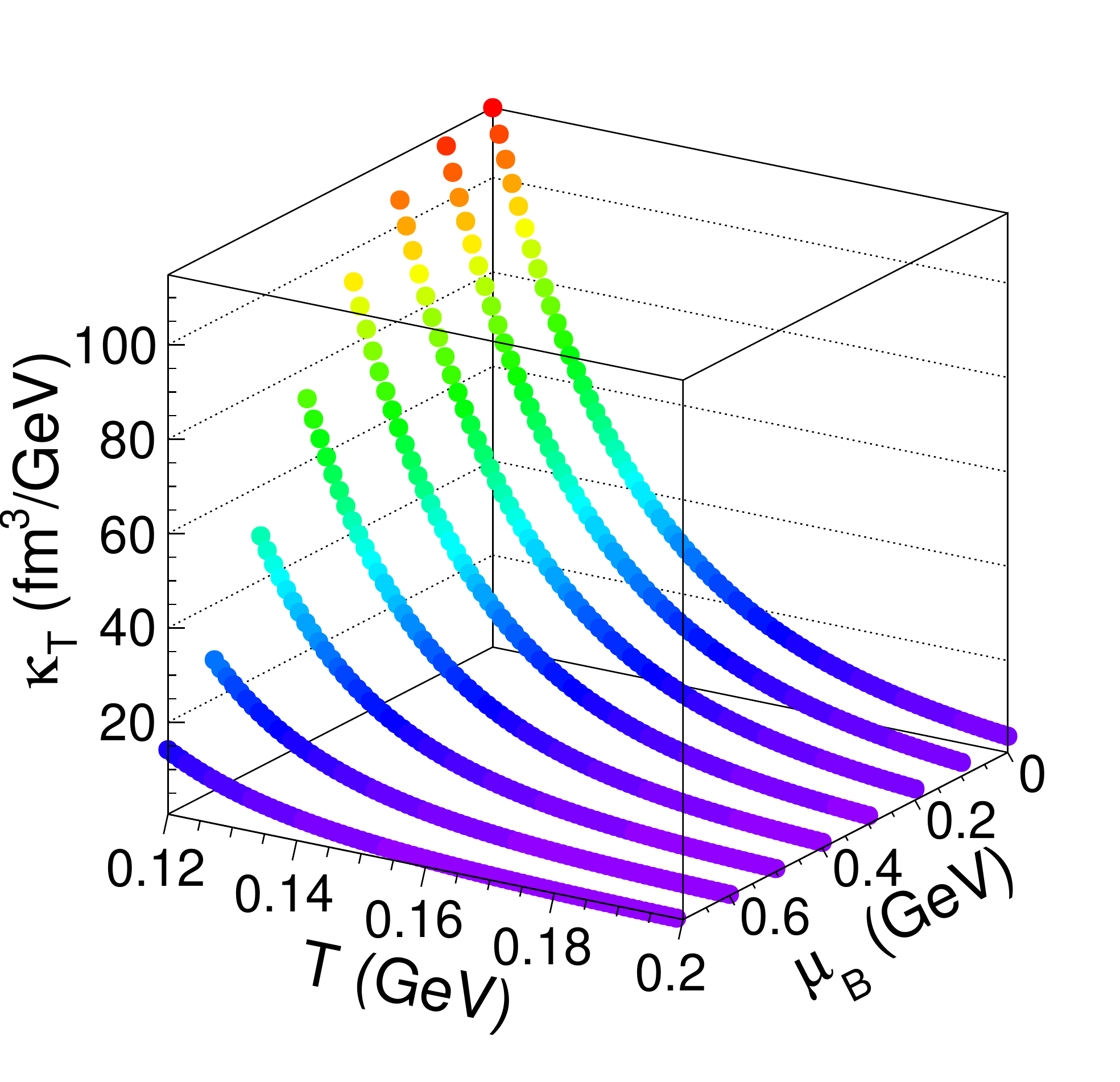}
\caption{(color online) Variation of $\kappa_T$ as a function of $T$ and $\mu_B$ in HRG model.}
\label{HRG3D}       
\end{figure}

Figure~\ref{kTmu} depicts the isothermal compressibility calculated in HRG model as a function of temperature for various $\mu_B$ with mass cut-off of 2 GeV. We notice that, $\kappa_T$ decreases with the temperature and has lower values for higher baryon chemical potential, which goes inline with the concept that as the number of baryon states increase, the system becomes incompressible. The effect of baryon chemical potential on $\kappa_T$ is more pronounced at a lower temperature.  

In Fig.~\ref{HRG3D}, a 3-dimensional (3D) plot of isothermal compressibility calculated in HRG model as a function of $T$ and $\mu_B$ is shown. It is observed that $\kappa_T$ decreases with $T$ at a constant $\mu_B$ and its values decrease with the increasing $\mu_B$ for a given temperature of the system.

\begin{figure}[ht!]
\includegraphics[width=0.45\textwidth]{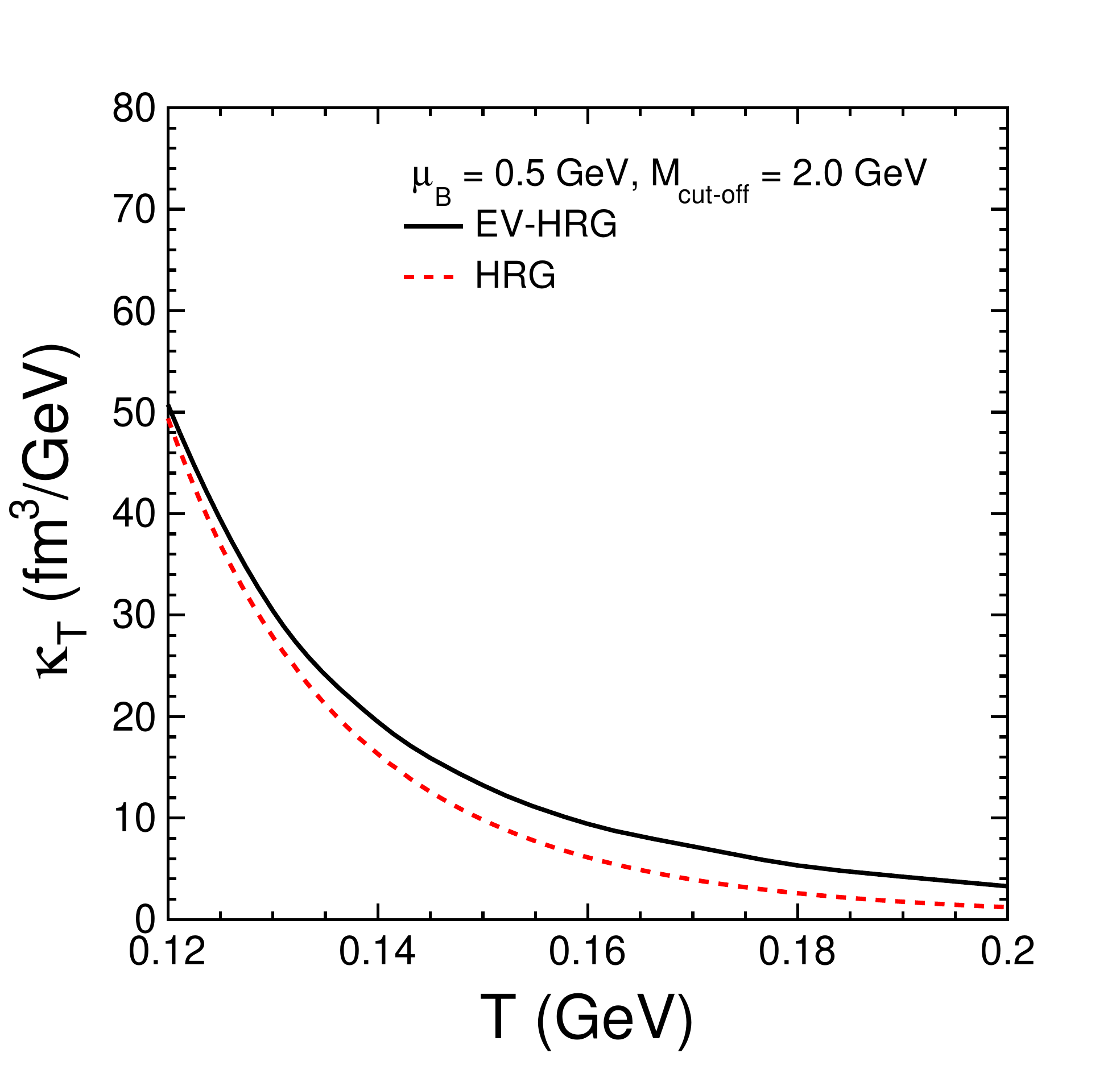}
\caption{(color online) $\kappa_T$ as a function of temperature. Black line is the result obtained in EV-HRG and dotted red line is from HRG calculations.}
\label{EVHRG}       
\end{figure}

Now, we want to compare the results of HRG and EV-HRG models. For this, we compare $\kappa_T$ calculated in both the models for $\mu_B$ = 0.5 GeV with mass cut-off = 2 GeV as shown in Fig.~\ref{EVHRG}. We notice that when $\kappa_T$ is varied as a function of $T$, the EV-HRG results lie above those obtained in HRG model and the difference is larger at a higher $T$. This finding tells that excluded-volume correction in ideal hadron gas model plays an important role as we go towards higher temperatures.

\begin{figure}[ht!]
\includegraphics[width=0.45\textwidth]{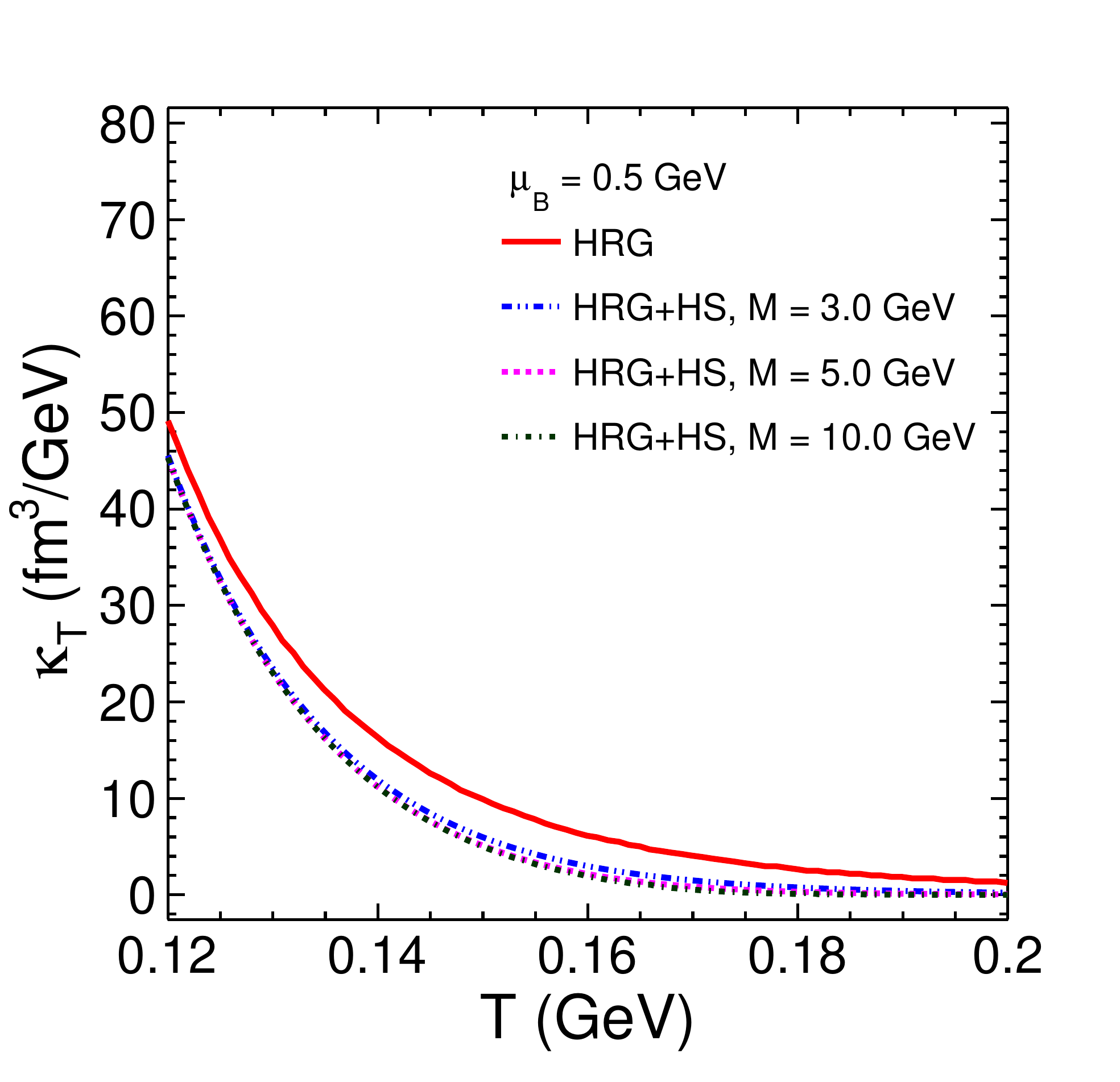}
\caption{(color online) $\kappa_T$ as a function of temperature in HRG model and HRG with Hagedorn mass spectrum for various mass cut-offs at $\mu_B$ = 0.5 GeV.}
\label{kTHagmass}       
\end{figure}

Figure~\ref{kTHagmass} represents the results of $\kappa_T$ estimated in the HRG model for $\mu_B$ = 0.5 GeV. We have also shown the results after adding the Hagedorn states (HS) in the grand partition function of HRG model for various upper mass limits.  Here, we take continuous mass spectrum in the calculation above hadron mass of 2 GeV. An upper mass limit in the mass integration is taken upto 10 GeV. The decrease in compressibility is found with the addition of higher HS into the system. The effect of Hagedorn states is prominent when the mass cut-off increases from 2 GeV to 3 GeV.

\begin{figure}[ht!]
\includegraphics[width=0.45\textwidth]{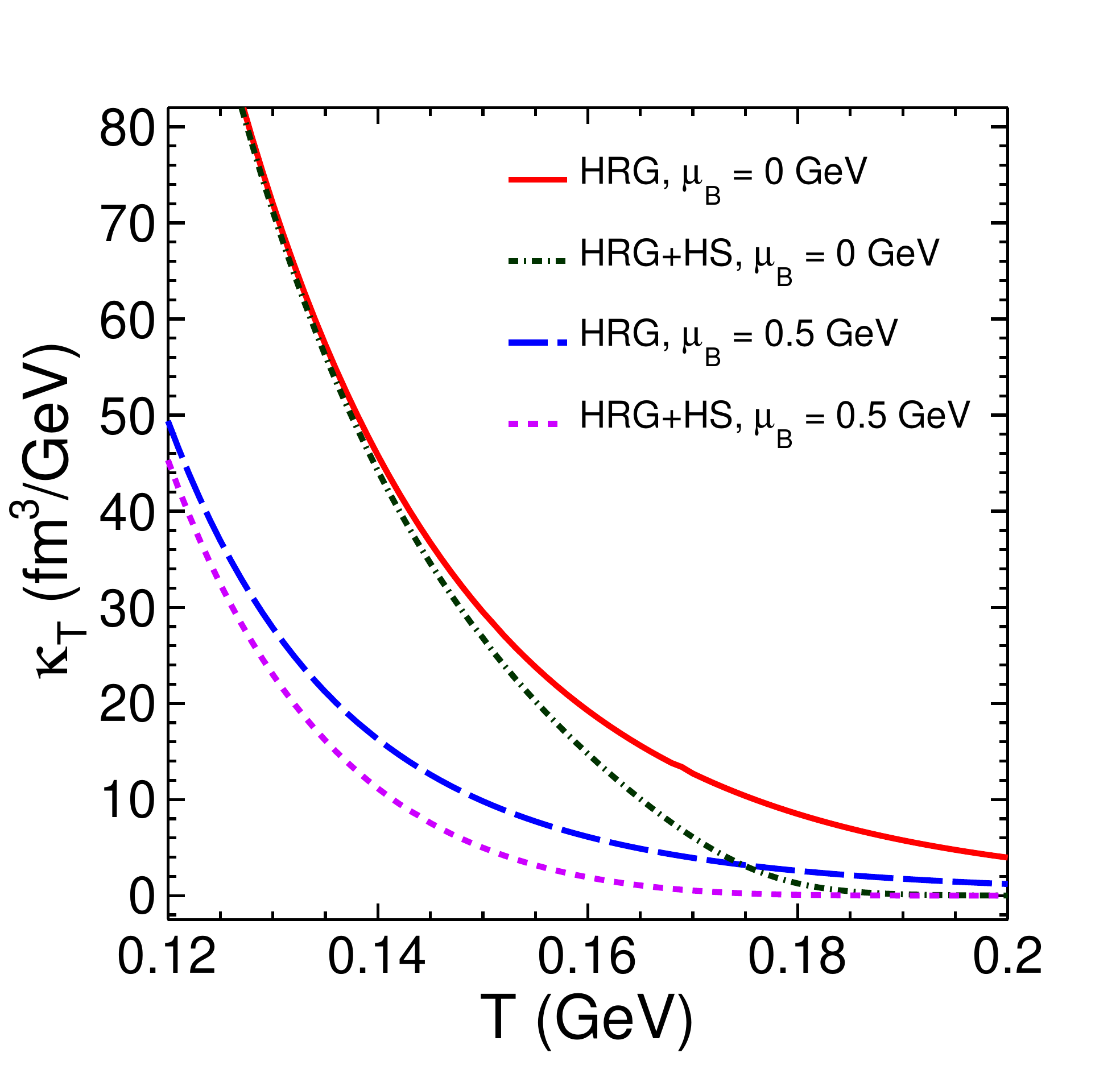}
\caption{(color online) $\kappa_T$ as a function of temperature in HRG model and HRG with Hagedorn mass spectrum for various $\mu_B$ at mass cut-off of 10 GeV.}  
\label{kTHagmu}      
\end{figure}

Figure~\ref{kTHagmu} depicts the impact of baryochemical potential on the isothermal compressibility for hadronic matter calculated in both HRG and HRG+HS models, where an upper mass limit of 2 and 10 GeV are taken in the HRG and HRG+HS models, respectively. It is evident from the figure that $\kappa_T$ has lower values particularly at a higher $T$ in case of HRG+HS in comparison to that obtained in HRG. Again, we notice that it decreases with the increasing baryonic chemical potential for both the cases. Addition of Hagedorn states reduce the values of $\kappa_T$ supports the earlier findings that inclusion of higher masses makes the system incompressible.   

\begin{figure}[ht!]
\includegraphics[width=0.50\textwidth]{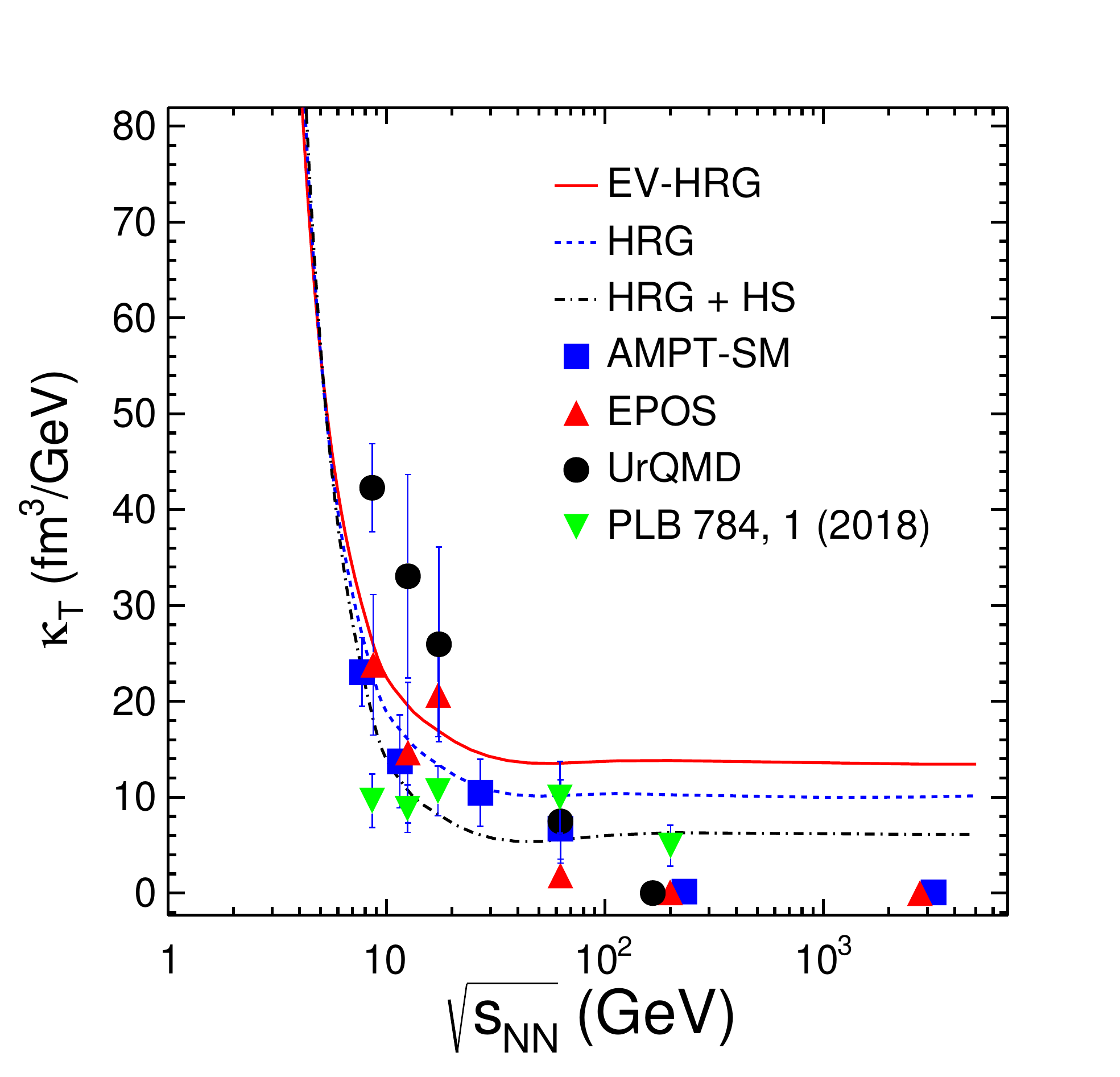}
\caption{(color online) The centre-of-mass energy dependence of $\kappa_T$ obtained in  HRG, excluded-volume HRG and HRG with Hagedorn states. 
Transport model results from AMPT, UrQMD and EPOS are shown and results from Ref.~\cite{Mukherjee:2017elm} are superimposed for comparison.}
\label{kTsNN}       
\end{figure}

In order to study the collision energy dependence of isothermal compressibility, we need a parametrisation of temperature and baryon chemical potential in terms of centre-of-mass energy~\cite{Cleymans:2005xv}: 
\begin{equation}
\mu_B=\frac{a}{1+b\sqrt{s_{NN}}},
\end{equation}
and,
\begin{equation}
T=c-d\mu_B^2-e\mu_B^4, 
\end{equation}
where the parameters in EV-HRG are: $a=1.482\pm0.0037$ GeV, $b=0.3517\pm0.009$ {GeV}$^{-1}$, $c=0.163\pm0.0021$ GeV, $d=0.170\pm0.02$ {GeV}$^{-1}$ and $ e=0.015\pm0.01$ {GeV}$^{-3}$. In case of HRG, the parameters are: $a=1.22\pm0.04$ GeV, $b=0.242\pm0.017$ {GeV}$^{-1}$, $c=0.170\pm0.003$GeV$, d=0.190\pm0.039$ {GeV}$^{-1}$ and $ e=0.0108\pm0.0074$ {GeV}$^{-3}$~\cite{Tiwari:2012}. Figure~\ref{kTsNN} represents $\kappa_T$ as a function of centre-of-mass energy ($\sqrt{s_{NN}}$) evaluated at the chemical freeze-out temperature ($T_{ch}$). In the course of evolution of a hot fireball created in heavy-ion collisions, the particle multiplicity is frozen at the chemical freeze-out, where the inelastic collisions cease. This chemical freeze-out surface thus corresponds to a fixed temperature and baryon chemical potential at a given collision energy. In order to estimate the isothermal compressibility at a given collision energy in the context of HRG, EV-HRG and HRG+HS, we use the chemical freeze-out parameters as the inputs.
 
  The solid line is the result obtained in EV-HRG model while the dashed line is the result obtained in HRG model. The dash-dotted line is the calculation done in the framework of a HRG model with the Hagedorn mass spectrum. We perceive that isothermal compressibility initially decreases rapidly with $\sqrt{s_{NN}}$ and gets saturated around collision energy of 10-20 GeV. This change of behaviour of $\kappa_T$ emphasises a very different nature of the system at lower collision energies in comparison to higher energies. The energy window at which such a change of behaviour of compressibility is observed could be the onset of phase transition from hadron gas to QGP. Significant differences are found between EV-HRG, HRG and HRG+HS results for $\sqrt{s_{NN}} >$ 10 GeV.  Further, we find that addition of Hagedorn states in the HRG model lowers down the values of isothermal compressibility. The results in HRG+HS framework are found to be closer to the results obtained in ref~\cite{Mukherjee:2017elm}, which are shown by the green inverted triangles. Other theoretical models, however, show some degree of deviation. This necessitates the inclusion of such Hagedorn mass spectrum while studying thermodynamical observables. We know that $\kappa_T$ is related to the particle multiplicity fluctuation, temperature and volume of the system formed in heavy-ion collisions \cite{Mukherjee:2017elm}. Multiplicity fluctuations are extracted from the event-by-event distributions of charged particle multiplicities in narrow centrality bins.  The dynamical part of the multiplicity fluctuations are obtained by removing the contributions from the number of participating nucleons. In Ref. \cite{Mukherjee:2017elm}, $\kappa_T$ is evaluated at various collision energies using the dynamical part of multiplicity fluctuations. In an experimental scenario, the chemical freeze-out temperatures are obtained by particle yields and their ratios, which are in agreement with those obtained from combined fits of net-charge, net-kaon and net-proton fluctuations ~\cite{Bellwied:2018tkc,Alba:2014eba}. Hence, it is reasonable to use the chemical freeze-out temperature as a proxy to the temperature where fluctuations happen to cease. We therefore use chemical freeze-out temperature for the estimation of isothermal compressibility.
In a similar fashion, the values of $\kappa_T$  have been estimated from three different transport models, such as, AMPT with string melting scenario (AMPT-SM), UrQMD and EPOS. These values have been plotted in Fig. \ref{kTsNN}. This study helps in confining the search for critical point in the QCD phase diagram to a centre-of-mass energy of 10-20 GeV per nucleon. It is observed that the matter created in heavy-ion collisions at the lower collision energies like in future facilities e.g. CBM@FAIR, NICA, Dubna is more compressible, unlike that created at the RHIC and LHC energies.

\section{Summary and Conclusions}
\label{summary}
In the present work, we have estimated the isothermal compressibility for hot and dense hadron gas using HRG and EV-HRG models. Here, hadrons and their resonances of masses only upto 2 GeV are considered in the system. We study the mass as well as baryon chemical potential dependence of $\kappa_T$ as a function of temperature. We have also shown the effect of Hagedorn mass spectrum on $\kappa_T$ after including the continuous mass spectra in the grand canonical partition function for hadron gas in HRG model. The lower limit of mass cut-off is set to 2 GeV in the mass integration of Hagedorn mass spectrum while the upper mass cut-offs vary from 3 -- 10 GeV. The findings of this study are summarised below:

\begin{itemize} 

\item Isothermal compressibility decreases with the increasing temperature which suggest that the system at higher temperature behaves as incompressible matter.

\item It is also observed that isothermal compressibility becomes lower for higher mass cut-offs and larger baryochemical potential. This again emphasises that, it is difficult to compress the matter with the addition of more number of baryons and heavier resonances. The effect of baryochemical potential is more prominent at lower temperatures.

\item The effect of continuous mass spectra is seen on $\kappa_T$, where heavier resonances having masses above 2 GeV are included into the system. Addition of more number of resonances lowers the values of isothermal compressibility.       

\item A collision energy dependence of $\kappa_T$ is studied in EV-HRG, HRG and HRG+HS models. Isothermal compressibility initially decreases rapidly with collision energy and it becomes saturated beyond $\sqrt{s_{NN}}$ = 10-20 GeV. This suggests a very different nature of the system formed in lower collision energies. Significant differences in the results from EV-HRG, HRG and HRG+HS are found for $\sqrt{s_{NN}} >$10 GeV. We emphasise the need to have more experimental data as a function of collision energy from very low to high energies.

\item Addition of Hagedorn mass spectrum in the HRG model reduces the isothermal compressibility and hence it approaches closer to the results obtained using experimental data at higher collision energies in comparison to the other theoretical calculations. This signifies the use of Hagedorn mass spectrum into the system while studying the isothermal compressibility.

\item We observe that the produced hadronic matter is more compressible at FAIR and NICA collision energies as compared to RHIC and LHC energies.

\end{itemize}

\section*{Acknowledgements}
The authors acknowledge the financial supports from ALICE Project No. SR/MF/PS-01/2014-IITI(G) of Department of Science \& Technology, Government of India.

\end{document}